\documentclass[submission,copyright,creativecommons]{eptcs}
\providecommand{\event}{PROLE 2015}

\usepackage{amsmath}
\usepackage{marginnote} % instead of the ACM forbidden \marginpar
\usepackage{color}
\usepackage{graphicx}
\usepackage{dsfont}   % for natural domains (N, B, ...)
\usepackage{algorithm,algorithmic}
\usepackage{listings}
\usepackage{multicol}
\usepackage[style=math]{k}
\usepackage{semantic}

\newtheorem{example}{Example}
\newcommand{\cellface}[1]{{\color{white!50!black}\sf #1}}

\newcommand{\heap}{\cellface{heap}}

\newcommand{\env}{\cellface{env}}

\newcommand{\kcompcell}{\cellface{k}}

\newcommand{\pathcondcell}{\cellface{path-condition}}
\newcommand{\mempc}{\cellface{init-struct}} %mem-cond
\newcommand{\stack}{\cellface{stack}}
\newcommand{\cfg}{cfg}

\newcommand{\KingSpec}{\mbox{\sc KindSpec 2.0}}

\newcommand{\maude}{\mbox{Maude}}
\newcommand{\curry}{\mbox{Curry}}
\newcommand{\haskell}{\mbox{Haskell}}

\newcommand{\C}{\mbox{\sc C}}
\newcommand{\KernelC}{\mbox{\sc KernelC}}

\newcommand{\Maude}{\mbox{\sc Maude}}

\newcommand{\SE}{\mbox{\sc SE}}
\newcommand{\Se}{\mbox{\textsc{Se}}}

\newcommand{\QuickSpec}{\mbox{\sc QuickSpec}}
\newcommand{\AbsSpec}{\mbox{\sc AbsSpec}}

\newcommand{\prgvar}[1]{\text{\tt {#1}}}
\newcommand{\symb}[1]{\text{\sf {#1}}}
\newcommand{\smalloperator}[1]{\text{$\operatorname{#1}$}}

\newcommand{\mapsTo}[2]{#1\mapsto{}#2}
\newcommand{\ie}{i.e.,}

\newcommand{\explain}[1]{\mathop{explain}(#1)}

\newcommand{\methods}{\mathcal{F}}
\newcommand{\obsM}{\mathcal{O}}
\newcommand{\modM}{\mathcal{M}}
\newcommand{\KernelCConfiguration}[0]{
\[
\kallLarge{cfg}{
  \kall{k}{\operatorname{K}}
  \kall{env}{\operatorname{Map}}
  \kall{stack}{\operatorname{List}}
  \kall{heap}{\operatorname{Map}}
}
\]%
}

\newcommand{\sePostCondIfTwo}[0]{
$$
\kallLarge{cfg}{
\ellipses{}
\\
\kmiddle{env}{
   \mapsTo{{\tt list}}{\symb{list}},
   \mapsTo{{\tt final}}{\symb{list}}
}
\\
\kallLarge{heap}{
    \mapsTo{\symb{list}}{(
    	\mapsTo{\prgvar{data}}{\operatorname{undef}},
	\mapsTo{\prgvar{prev}}{\operatorname{undef}},
	\mapsTo{\prgvar{next}}{\symb{list.next}}
    )}
    \\
    \mapsTo{\symb{list.next}}{\operatorname{undef}}
}
\\
\ellipses{}
}
$$
}

\usepackage{url}

\begin{document}

%\title{Automatic Inference of Specifications using the K framework
\title{Automatic Inference of Specifications in the \K\ Framework \thanks{This work has been partially supported by the EU (FEDER)
and Spanish MINECO under grants
TIN2015-69175-C4-1-R
and TIN2013-45732-C4-1-P, and
by Generalitat Valenciana ref. PROMETEOII/2015/013.}\\
%{\em Work in Progress}
}

\author{Mar\'{\i}a Alpuente \qquad Daniel Pardo \qquad Alicia Villanueva
\institute{DSIC,  Universitat Polit\`ecnica de Val\`encia\\ Camino de Vera s/n \\ 46022 Valencia, Spain}
% \institute{School of Computer Science and Engineering\\
% University of New South Wales\thanks{A fine university.}\\
% Sydney, Australia}
\email{\{alpuente,daparpon,villanue\}@dsic.upv.es}}

\def\titlerunning{Automatic Inference of Specifications in the \K\ Framework}
\def\authorrunning{Mar\'{\i}a Alpuente, Daniel Pardo, Alicia Villanueva}

\maketitle

\begin{abstract}
%  Formal specifications can be used for various software engineering
%  activities ranging from finding errors to documenting software and
Despite its many unquestionable benefits, formal specifications are not widely used in industrial software development.
In order to reduce the time and effort required to write formal specifications, in this paper we propose
a technique for automatically discovering
  specifications from real code. 
%  for heap-manipulating programs is a challenging
 % task. 
The proposed methodology relies on the  symbolic execution capabilities recently provided by the \K\ framework that we exploit to  automatically
  infer  formal specifications from programs that are written in a non--trivial  fragment of \C,\ called \KernelC.
Roughly speaking, our symbolic analysis of \KernelC\ programs
explains the %(symbolic) 
execution of a  (modifier) function   by
using other (observer) routines in the program.
  % which is based on the symbolic execution 
%
  We implemented our technique in the automated tool \KingSpec,\ which generates axioms that
  describe the precise input/output behavior of \C\ routines that
  handle pointer-based structures (i.e.,\ result values and state
  change).    We describe the implementation of our system and discuss the differences w.r.t.\ our previous
  work on inferring specifications from \C\ code.
% These specifications can be written either in \MLlong\

%  Although the final goal of this work is the same of our previous
%  work, \ie{} inferring specifications for \C\ code, there are some
%  significant differences in the strategy we follow to accomplish.
%% These specifications can be written either in \MLlong\
%   itself, which is useful for further automated analysis within the
%   \K\ formal environment, or in sugared axiomatic form, which favors
%   better human inspection.
  % Since we rely on rewriting logic \K\
  % semantics specification of programming languages, our approach can
  % be easily extended to any language for which %that
  % a formal semantics in \K\ is given.

\end{abstract}

%\keywords 
%Specifications inference, Symbolic execution, Matching Logic

%----------------------------------SECTION-----------------------------------%
\section{Introduction}\label{sec:introduction}
% !TEX root = ./PROLE2015-APV.tex
%Formal specifications can document code unambiguously and are  important for rigorous software development. 
Formal specifications can be used for various software engineering
  activities ranging from documenting software to automated debugging, verification, and  test-case generation.
  However, there are a variety of reasons why software companies do not currently consider formal specification to be cost-effective to apply;
 these include time, complexity, and tool support.
 Specification inference can help to mitigate these problems and is also useful for legacy program understanding and malware deobfuscation
%, where the challenge is to understand what the malicious code is doing
 \cite{ChristodorescuJK07}.  
%Algorithm verification and program testing can often diagnose discrepancies between implementation and specification automatically.
%Unfortunately, formal specifications are notoriously hard to write and debug, and many programs lack appropriate documentation or it may be too low-level to understand.  

This paper describes our ongoing work in developing a specification inference system for heap-manipulating programs that are written in a non-trivial fragment of \C\ called \KernelC\ 
%This paper describes a rule-based technique that automatically infers high-level, formal specifications from  heap-manipulating code  that is written in
% \KernelC\ 
 \cite{rosu-schulte-serbanuta-2009-rv}, which includes functions, structures, pointers, and I/O primitives.
%
% by using the notation of \MLlong\ (\ML), which is a novel program verification foundation that is built upon operational semantics \cite{RosuS2011}. 
%
 We rely on the (rewriting logic) semantic framework \K\ \cite{Rosu2015}, which
 facilitates the development  of executable semantics of programming languages
 and also allows formal analysis tools for the defined languages to be derived with
 %reasonable 
 minimal effort. 
%

% In \K, programming languages can be defined using
%  configurations, computations and rules.  
%, called configuration {\em patterns}. 
 
 A language definition in \K essentially consists of three parts: the BNF
 language syntax (annotated with \K\-specific attributes), the
 structure of program configurations, and the semantic rules.
 Similarly to the classic operational semantics, program
 configurations contain an encoding for the environment, the heap,
 stacks, etc. and are represented as algebraic datatypes in
 \K. Program configurations organize the state in units called {\em
   cells}, which are labeled and can be nested.

For example, following the \K{} notation, the program configuration
\begin{equation}\label{eq:config}
%\[
\kallLarge{cfg}{
	\kall{k}{\operatorname{tv}(int,0)}
	\kall{env}{\mapsTo{\prgvar{x}}{\symb{x}}}
	\kall{heap}{\mapsTo{\symb{x}}{\operatorname{tv}(int,5)}}
}
%\]
\end{equation} 
%\mapsTo{\symb{s}}({\mapsTo{{\tt s.size}}{4}}, 
models %a configuration corresponding to
the final 
state of a computation whose return value is the integer 0
(stored in the \kcompcell{} cell, which contains the current code to be
run), while
program variable \prgvar{x} (stored in the \env{} cell)
has the %integer 
value 5 (stored in the memory
address given by \symb{x} in
the \heap{} cell, %. The \heap{} cell stores
where information about pointers and data
structures is recorded). %\footnote{
Variables representing symbolic memory addresses are
  written in sans-serif font. 

%Also in the heap{} cell, the field {\tt size} of
%the data structure object \symb{s} is 4.
%
In \K,\ the configuration (\ref{eq:config}) is a friendly representation for the   %algebraic 
term 
\begin{center}
\begin{verbatim}
<cfg>
       <k> tv(int,0) </k>
       <env> x => pointer(x) </env>
       <heap> pointer(x) => tv(int,5) </heap>
</cfg>
\end{verbatim}
\end{center}
%pointer(s) => (size => 4,
%
Symbolic execution (SE) is a well-known program analysis technique
that allows the program to be executed using {\em symbolic} input
values instead of actual (concrete) data so that it executes the
program by manipulating program expressions involving the symbolic
values \cite{King1976,PasareanuV09}.  Unlike concrete execution, where
the path taken is determined by the input, in symbolic execution the
program can take any feasible path. That path is given by a logical
constraint on %the input symbols,
past and present values of the variables, called {\em path
  condition} because it is formed by constraints that are accumulated
  on the path taken by the execution to reach the current program
point.  Each symbolic execution path stands for many actual program
runs (in fact, for exactly the set of runs whose concrete values satisfy
the logical constraints).  One of the traditional drawbacks of SE-based techniques is the high cost of decision procedures to
solve path conditions.  Recently, SE has found renewed interest due in
part to the huge recent advances in decision procedures for logical
satisfiability.

\K\ semantics is traditionally\footnote{\K's backend is currently being ported into Java, and \K{} 4.0 is expected to be released when the Java backend is deemed a suitable complete replacement for Maude.}  
 compiled into Maude \cite{maude-book} for execution, debugging, and model checking.
\K\ implements reachability logic in the same way that Maude implements rewriting logic. 
In reachability logic, a particular class of first-order formulas
with equality (%are used that are      
encoded as
(boolean) terms with logical variables and constraints over them) is used. These formulas, called {\em patterns},  specify those concrete configurations that match the pattern algebraic
structure and satisfy its constraints.
Since patterns allow  logical variables and constraints over them, by
using patterns,  \K\ rewriting becomes {\em symbolic execution} %(SE)
with the semantic rules of the language
\cite{Arusoaie-SymbolicK2014}. The SMT solver {\sf Z3} \cite{Z3-2008}
is used in \K\ for %{\color{red}simplifying and satisfiability} checking the path conditions.
%simplifying and 
checking the satisfiability of the path constraints.
%
%

%A framework has been defined in \cite{Arusoaie-SymbolicK2014}   that
%effectively supports the symbolic execution of \K{} specifications.

Symbolic execution in \K{}  relies on an automated transformation of both \K{} configurations and \K{} rules
into corresponding symbolic \K{} configurations (i.e.,\ patterns) and symbolic \K{} rules that capture all required symbolic ingredients:
symbolic values for data structure fields and program variables; path conditions that constrain the variables in  cells;
multiple branches when a condition is reached during execution, etc.
%\alicia{yo bajar\'{\i}a aqu\'{\i} la ``definici\'on'' de pattern:\\
%We call \emph{pattern} to these symbolic configurations, which can be
%seen as a particular class of first-order formulas  with equality, encoded as
%(boolean) terms with logical variables and constraints over them. % Patterns model those concrete configurations  matching the pattern algebraic
%% structure and satisfying ??its constraints??.
%Since patterns allow  logical variables and constraints over these variables, by
%using patterns  \K\ rewriting becomes {\em symbolic execution} (SE)
%with the semantic rules of the language
%\cite{Arusoaie-SymbolicK2014}.}
%A  symbolic configuration 
% that is formed by a \K{} configuration 
%(boolean terms with variables) 
%together with a path condition that constraints %over 
%the boolean terms in the \K{} configuration. 
The transformed, symbolic rules define how symbolic configurations  are
rewritten during computation. Roughly speaking,  each data structure field and program variable   originally holds  an initial,  symbolic value.
%
%Symbolic rules define how symbolic configurations are
%transformed during computation where a 
Then, by symbolically executing  %in \K{}  
 % in this framework means that
% each data structure field and program variable initially holds a
%symbolic value.  
%Then, 
a program statement, the
configuration cells (such as 
%$\mathit{env}$, $\mathit{mem}$ and $\mathit{cfg}$ 
\kcompcell{},  \env{} and  \heap{} in the example above)
 are  updated by mapping fields and variables
to  new symbolic values that are represented as symbolic expressions, while
the   path conditions (stored in the \pathcondcell\ cell) are   correspondingly updated at each branching
point. 

%\alicia{Si quitamos los ejemplos arriba, este tambi\'en...\\

%\maria{(revisar)\\
For instance, the following  pattern 
\[
\kallLarge{cfg}{\footnotesize
 \kall{k}{
   \operatorname{tv}(int,0)
 }\\
  \kmiddle{env}{
%    \mapsTo{\prgvar{s}}{\symb{s}},
    \mapsTo{\prgvar{x}}{\symb{x}},\mapsTo{\prgvar{s}}{\symb{s}}
  }\\[-.5ex]
  \kmiddle{heap}{\footnotesize
    \mapsTo{\symb{s}}{(\mapsTo{\prgvar{size}}{\symb{?s.size}}, \mapsTo{\prgvar{capacity}}{\symb{?s.capacity}})}}
 }
\kallLarge{\text{\pathcondcell{}}}{\footnotesize\symb{s} \neq \symb{NULL} \wedge
\symb{?s.size} > 0}%\geq \symb{?s.capacity}}
\]
%
%   For example, the
% pattern
% $$\kall{cfg}{\kall{env}{\ellipses{}\mapsTo{\prgvar{top}}{\symb{?top}}\ellipses{}}\kall{heap}{\ellipses{}\operatorname{list}(\symb{?top})(\symb{L})\ellipses{}}}\wedge~\mbox{$\symb{L}\neq{}\operatorname{empty}$}$$
specifies the set of configurations as follows: (1) the \kcompcell{}
  cell contains the integer value 0; (2) in the \env{} cell, program variable {\tt x} (in typographic font) 
is associated to the memory address \symb{x} and 
%object 
{\tt s} %of a \C{} structured type 
is bound to the pointer \symb{s}; %(sans-serif font),
%program variable {\tt size} has the symbolic value in the size field of symbolic expression {\sf s} 
 and (3) in the \heap{} cell, the field \symb{size} of 
 \symb{s} contains the symbolic value \symb{?s.size}
 (symbolic values are preceded by a question mark). Additionally,
 \symb{s} is not null and the value of its size field is greater than 0.

%\alicia{he simplificado la \'ultima constraint y quiz\'as podr\'{\i}a
%  incluso quitarse el ``capacity'' pero de momento lo dejo as\'{\i}}
% or equal to its capacity field. 
%Hence, separation (meaning that the heap can be split into two disjoint parts where the separate  formulas hold \cite{Reynolds2002}), is achieved at the structural (i.e.,\ term) level.
%Marked variables like \symb{?top} are bound (i.e.,\ existentially quantified) over the pattern, while \symb{L} is free.
%

%Patterns can be interpreted as a particular
%class of first-order formulas with equality: a pattern specifies those
%configurations that match its algebraic structure and satisfy its
%constraints. 
%
%\alicia{eliminar? \\
%
%By allowing specifications to directly refer to the structure of the
%program configurations, \K\ facilitates access and reasoning about
%rich sub-patterns of the program states, such as disjoint lists, trees
%and graphs, or any shared mutable data structures that are dynamically
%allocated in the heap.
%}

In this paper, we redesign the technique of \cite{AFV13} for discovering %formal
specifications for heap-manipulating programs by adapting the symbolic
infrastructure of \K\
%which is driven by \ML\ formulas.
to support the specification inference process for \KernelC\ programs. Specification inference is the task of discovering high-level specifications that closely describe the program behavior.
%Obviously, these specifications can only be correct with respect to user intent if the original program is correct itself.
%But even if it is not correct, the ascertained specification can still be very helpful in several important scenarios such as improving program understanding, synthesizing test units, and helping the programmer to debug the code.
%
Given a program {\tt P}, the specification discovery problem for {\tt P} is typically described as the problem of inferring a likely specification for 
every function {\tt m} in {\tt P} that uses I/O primitives and/or modifies 
the state of encapsulated, dynamic data structures defined in the program.  
Following the standard terminology, any such function {\tt  m} is called a \emph{modifier}.  
The intended specification for {\tt   m} is to be cleanly expressed by using any combination of the non-modifier functions of {\tt P} (\ie\ functions, called \emph{observers}), which inspect the program state and return 
values expressing some information about the encapsulated data.
However, because  the \C{} language does not enforce   data encapsulation, 
we cannot assume purity of any function: 
every function in the program can potentially change the execution state, including the \heap{} component of the state.
In other words, any function  can potentially be a \emph{modifier}; hence we simply define an \emph{observer} as any function whose return type is different from {\tt void} (\ie{} potentially expresses
a property   concerning the final  \emph{heap} contents or the return value of the function call). %arguments . 
%

%%Movido este parrafo a la seccion 3:
%The proposed inference technique relies on the classification scheme
%developed in \cite{Liskov86} for data abstractions in general, where a
%function (method) may be either a {\it constructor}, {\it modifier} or
%{\it observer}.  A constructor returns a new object of the class from
%scratch (i.e., without taking the object as an input parameter).  A
%modifier alters an existing class instance (i.e., it changes the state
%of one or more of the data attributes in the instance).  An observer
%inspects the object and returns a value characterizing one or more of
%its state attributes.  We do not assume the traditional premise of the
%original classification in \cite{Liskov86} that states that observer
%functions do not cause side effects on the state.  This is because we
%want to apply our technique to any program, % that
%maybe written by third-party software producers that may not follow
%the observer purity discipline.

%Our symbolic analysis of \KernelC\ programs
%explains the (symbolic) execution of a \emph{modifier} function $m$ by
%using other (observer) routines in the program. 
The key idea
behind our inference methodology  was originally described in \cite{AFV13}.  % We have developed a
% procedure %on top of \MatchC\
% that is executed symbolically in \
% K.
%
Given a \emph{modifier} procedure for which we desire to obtain a
specification, 
we start from an initial  symbolic state $s$ and 
symbolically evaluate  $m$ on $s$ to obtain as a result a set of pairs
$(s,s')$ of 
 initial and final symbolic states,
respectively. 
%its symbolic execution in   \K{} %specification
%delivers an environment for the \emph{post-state} that gives
% the (symbolic) return value for
%each program variable and field in terms of the values of program variables
%and fields in the symbolic \emph{pre-state}. 
Then, the observer methods   in
the program are used to explain the computed %post-states. 
 final  symbolic states. % $s'$. 
 This is
achieved by analyzing the results of the symbolic execution of each
observer method $o$ when it is fed with (suitable information that is 
easily  extracted from)  $s$ and $s'$. More precisely, for each pair
$(s,s')$ of %refined 
initial and final states, a pre/post statement is synthesized where
the precondition is expressed in terms of the observers that
\emph{explain} the initial state $s$, whereas the
postcondition contains the observers that \emph{explain} the final
state $s'$. To
express  a (partial) observational abstraction or
explanation for (the constraints in) a given state in terms of the observer $o$, our criterion  is   that $o$  computes the same symbolic values at the end of   all its  symbolic execution branches.
%the computed post-states. 
%%%% MOVER
% Following the common symbolic
% approach to finitize program execution \cite{TaghdiriJ07}, loops are
% handled by unrolling them to a fixed depth, i.e.,\ the loop is
% executed a fixed number of times.  We also limit the heap's size so
% that the space of possible heaps is also finitized.
%%%%

In contrast to \cite{AFV13}, in this work we rely on the newly defined
symbolic machinery for \K, while \cite{AFV13} was built on a symbolic infrastructure for \KernelC\
  that we manually developed in a quite ad-hoc and error prone way, by reusing some spare features of the formal verifier {\sf MatchC} \cite{RosuS11}. This strategic technological change will
allow us to define a generic and more robust framework for the inference of
specifications of languages defined within the \K\ framework. 
Also differently from \cite{AFV13}, here we fully  use  the 
lazy initialization approach of \cite{APV2009-LazyInit} in order to deal with
complex data structures and pointers, which were only partially
adopted in our previous work.
With lazy initialization, the first time an uninitialized field or reference is accessed,
instead of considering all the
possible instances of these data structures, 
 the execution is non-deterministically branched by simply initializing the field
to the different scenarios: the field is null, points to a new object
with uninitialized fields, or points to an already created object.
%
%\alicia{mover a la parte de inferencia y dejar solo lo de arriba mas breve?\\
%
% In order to compute suitable explanations for the
%routine $m$, we symbolically evaluate the observer methods on each
%state $s^{0}_{i}$ and $s^{f}_{i}$ so that when the observer returns
%

\paragraph*{Contributions}
We summarize the main contributions of this paper   as follows: 
\begin{itemize}
\item  In the current \K{} system, we revisit the approach to extract lightweight
  specifications from heap-manipulating code of \cite{AFV13}, which    consists of a
  symbolic analysis that explores and summarizes the behavior of a
  {\em modifier} %program 
routine by using other available routines in
  the program, called {\em observers}.
  
  This corresponds to the primary motivation for this work: to migrate
  the specification discovery technique of \cite{AFV13} to the
  holistic framework of the latest \K release, which is based on
  symbolic execution, whereas \cite{AFV13} relied on the {\sf MatchC}
  verification infrastructure of the old \K platform, which is
  currently unsupported. 
%  Moreover, our previous approach built on a symbolic infrastructure for \KernelC\
%   that we manually developed in a quite ad-hoc and error prone way, by reusing some spare features of the formal verifier {\sf MatchC} .
  %  This was a manual and an error-prone process 
%  Hence, we now we gain robustness not only thanks
%  to the portability and mainteinance of the new technology.
%  , but also
%  because the previous approach was develpe error-prone process), whereas in this work just the
%  specific part of the complex data manipulation had to be defined.}

\item  We   adapt  the symbolic mechanism of \K{} to deal with
  \KernelC, also adapting and implementing the lazy initialization
  technique for manipulating complex \KernelC\ input data.
\item  We   implement our  specification inference technique in the \KingSpec\
% that targets \KernelC\
system, which fully builds on the %
%takes advantage of the unsat core  generated by the 
   capabilities of the   SMT solver Z3~\cite{Z3-2008}   to not only prove the (accumulated path) constraints as in \K{}  but also to incrementally simplify  them on the fly.
  % coupled to \K. 

Moreover, the   synthesized  pre/post axioms %that abstract from any implementation details
are further
 simplified (to be given more compact representation) and are eventually presented in a more friendly sugared form
 that abstracts from any implementation details.

%
%
%\item  \alicia{(depends on our experiments)} We have applied our technique on 
%\KernelC\ implementations of \emph{collection} libraries   for manipulating linked lists in \C\  such as those 
%provided in the GDSL generic data structure library \cite{GDSL}.
%%
%To evaluate our method, we performed two experiments
%using the prototype tool implementation \KingSpec.\ In one experiment, we applied 
%it to a small component for which complete specifications 
%were already available: a standard implementation  of 
%sets in \C\ by using linked lists.  In the second experiment, we applied the tool to 
%other kinds of heap-allocated   mutable structures.
%These components are more typical of 
% object-oriented style code written in C. We wrote specifications for 
%the source code in the style of the specifications our tool 
%infers; this allowed us to assess the accuracy of our  generated specifications, which properly
%target the properties that constrain the structure of the data in the heap. 
%Actually, we found that only in a few cases there was  a significant loss of information. 
\end{itemize}

 \paragraph*{Related work} 
The wide interest in program specifications as helpers for other
analysis, validation, and verification processes have resulted in
numerous approaches for (semi-)automatic computation of different kinds of specifications.
%, from algebraic properties to the input-output behavior of the system
%The automatic generation of likely specifications 
% either in the form of contracts, interfaces, summaries, assumptions, invariants, properties, component abstractions, process models, rules, graphs, automatas, etc. 
 %from program code has received increasing attention. 
Specifications can be property oriented (i.e.,\ described by pre-/post conditions or functional code); stateful (i.e.,\ described by some form of state machine); or intensional (i.e.,\ described by axioms), and can take the form of contracts, interfaces, summaries, assumptions, invariants, properties, component abstractions, process models, rules, graphs, automata, etc. 
In this work, we focus on   input-output relations: 
%discover specifications for the relation of the
%so-called input-output behavior of the system: 
given a
 precondition for the state, we infer which modifications in the state  are
 implied, and we express the relations  as logical implications that reuse the program
 functions themselves, thus improving comprehension since the user is acquainted with them.  
A thorough comparison with
the related literature can be found in \cite{AFV13}. Here we only try to cover those lines of research that have influenced our work the most.  

%input-output relations.
%

%Our approach also differs from most of the above because we do not infer abstract properties by observations of (concrete) program runs.
Our axiomatic representation %of functions and of their effects  
is inspired by  \cite{TillmannCS06}, which relies 
 %However, our approach does not rely 
 on a model checker for symbolic execution and
 %, as opposed to Tillmann's approach.
% Also, we do not generate the output as 
 generates either Spec\# specifications or parameterized unit tests. In contrast to \cite{TillmannCS06}, we take advantage of \K\ symbolic capabilities to generate simpler and more accurate formulas that
avoid reasoning with the global heap because  the different pieces of the heap that are reachable from the  function argument addresses are kept separate. 
 Unlike our 
 symbolic approach, Daikon \cite{Daikon} 
and DIDUCE \cite{DIDUCE}  detect program invariants 
by extensive  testing. 
Also, Henkel and Diwan \cite{Diwan} %  built a tool that
dynamically
discover  specifications for interfaces of Java classes by
 %by  first generating, using the class signature, many test cases   that consist of terms representing sequences of method invocations, and then 
generalizing  the results of automated tests runs as  an algebraic specification.  
\QuickSpec\   \cite{CSH2010}  %is another inference tool that is based on testing
%and can be used to 
relies on the automated testing tool  {\sf QuickCheck}  to distill general  laws that  a \haskell{} program satisfies.
Whereas Daikon discovers invariants that hold 
at existing program points, \QuickSpec\ discovers equations between arbitrary terms that are constructed using an API, similarly
to \cite{Diwan}.  
% Also, they use a similar overall approach that is based on testing: they generate terms and evaluate them, then dynamically 
%identify terms that are equal, and finally generate equations, filtering away redundant ones.  
\AbsSpec{} \cite{BacciCFV12} is a semantic-based  inference method that  relies on abstract interpretation and generates  laws for \curry\ programs in the style of \QuickSpec.   
A different   abstract interpretation approach to infer approximate specifications is  \cite{TaghdiriJ07}.
A combination of symbolic execution with dynamic testing is used in Dysy \cite{dysy}. 
An alternative approach to software specification discovery is based on inductive matching learning: rather than using test cases to validate a tentative specification, they are used as examples to {\em induce} the specification (e.g.,\ \cite{WhaleyML02,GPasareanu}). Finally, Ghezzi \emph{et al.} \cite{Ghezzi09} infer specifications for container-like classes and express them as finite state automata that are supplemented  with
graph transformation rules.

This work improves existing approaches in the literature in several
ways. 
%Some of  the above proposals observe that conditional equations would be useful \cite{dysy, Ghezzi09}, but neither tool generates them nor   include associative or commutative operators, which 
Thanks to the handling of \Maude's (hence \K's) equational attributes \cite{maude-book}, algebraic laws such  as associativity, commutatitvity, or identity are  naturally %inferred and 
supported in our approach, %which brings
%By supporting the modular combination of associative, commutative, idempotent and unity  equational attributes for function symbols (which makes these combinations transparent to the developer), the \K\ framework 
%naturally conveys enough expressive power to  
which 1) leads to simpler and more efficient specifications, and 2) makes it easy to reason  about typed data structures such as   lists (list concatenation  is associative with identity  element $nil$), multisets (bag insertion  is associative-commutative with identity  $\emptyset$), and sets (set insertion is associative-commutative-idempotent  with identity  $\emptyset$). 
 As a further advantage w.r.t.\   \cite{TillmannCS06}, in our framework, the correctness of the 
% inferred axioms can be checked automatically  by using the %very same
%  \MatchC{} verifier.
% Also, our methodology
%can be easily applied to any language which is given a semantics in the \K\ framework.
%Moreover,  correctness of the 
delivered specifications can be automatically ensured by using the 
existing \K\ formal tools
\cite{Rosu2015} .
Since our approach is generic and not tied to   the \K\ semantics
specification of \KernelC,\ we expect the methodology developed in this work to be
  easily extendable to other languages for which a \K\ semantics
is given. %like C, Java, JavaScript and Python %Java 1.4, Scheme and Verilog

\paragraph*{Plan of the paper}

In Section \ref{prelim}, we summarize the key concepts of the \K\ framework that are crucial for this work. 
Section \ref{sec:runningExample} %\ref{sec:symbolic} 
 introduces a running example that is used as a case study   throughout the paper to discuss the adequacy %, performance, 
 and effectiveness of the proposed inference methodology.
%It also allows us to 
%outline the major research problems addressed. 
Section~\ref{sec:SEinK} presents how we had to adapt %pinpoints several technical difficulties we found %had to face when  applying
%when applying
 the symbolic machinery  of \K to support specification discovery.
% our inference approach.
 Finally, Section~\ref{sec:inference} describes %upgrades the inference algorithms of \cite{AFV13} 
our  specification inference procedure  % technique
% to the new \K\ symbolic platform %mechanize  %our  updated  inference approach, 
 %and provides some experimental results. 
%Finally, Section~\ref{sec:future}  
and
discusses  directions for future work. %the related work and concludes.

%\section{Preliminaries}
\section{The \K\ Framework}\label{prelim}
% !TEX root = ./PROLE2015-APV.tex

%\section{Preliminaries}
%\maria{Mencionar la \ML\ me parece 
%un poco obsoleto: los nuevos trabajos permiten razonar sobre la din\'amica del sistema con la Reachability Logic (que es una extensi\'on de la   \ML\ ). De hecho, ya he visto que t\'u tambi\'en  lo hab\'{\i}as quitado (la subsecci\'on entera) m\'as abajo.}

In this section, we recall the fundamental concepts 
of the \K\ semantic framework \cite{RosuS2010}. %and \MLlong{}.

%\subsection{The \K\ framework}

%\K{} is an executable semantic framework in which programming languages, calculi, as well as type systems and formal analysis tools can be defined making use of configurations, computations, and rules.
\K{} \cite{RosuS2010} is a framework for engineering language semantics. Given
a syntax and a semantics of a language, \K{} generates a parser, an
interpreter, and formal analysis tools such as model checkers
and deductive theorem provers at no additional cost. It also supports
various backends, such as \maude{} and, experimentally, Coq.
In other words, language semantics defined in \K{} can be translated into \maude{}
or Coq definitions. %The most c
Complete formal program semantics %in the literature 
for Scheme, Java 1.4, JavaScript, Python, Verilog, and C are currently available in \K \cite{Rosu2015,RosuS2010}.
%\K\ semantics are 
%compiled into Maude \cite{maude-book} for execution, debugging, and model checking.

Program configurations are represented in \K\ as potentially nested structures of labeled cells (or containers) that represent the program state.
They include a computation stack or continuation  (named \kcompcell{}), environments (\env{}, 
 \heap{}), and a call stack (\stack{}), among others.
\K{} cells %are containers that 
can be lists, maps, (multi)sets of computations,
or a multiset of other cells.  
Computations carry ``computational meaning'' as special nested list structures that sequentialize computational tasks, such as fragments of a program. 
The  part of the \K{} configuration structure for the \KernelC{}
semantics that is relevant to this work is shown below.
%
% \daniel{Mantenemos stack o lo quitamos? En la práctica se usa para controlar la recursión y las llamadas anidadas a funciones (call stack), pero en el documento no se usa para nada...}
%\alicia{Yo opto por mantenerlo, tampoco hace danyo y no nos quita espacio}
%
\KernelCConfiguration{}%
%Containers (or cells) in a configuration represent pieces of the program state,
%including a computation stack or continuation  (named \kcompcell{}), environments (\env{}, 
% \heap{}), and a call stack (\stack{}), among others.

Rules in \K\ state how configurations (terms) %can 
evolve throughout the computation. 
Similarly to configurations, rules can also be graphically represented and are split in two levels. % and state how configurations change. 
Changes in the current configuration (which is shown in the upper level) are explicitly represented by underlining the part of the configuration that changes. 
The new value that substitutes the one that changes is written below the underlined part.

%\maria{Unificar el font que se usa en la regla azucarada con la
%  notacion azucarada que se usa  en la configuracion  K  de la
%  introducci\'on.}\\
%\alicia{Es que esto es un patr\'on, son variables a nivel de
%  definici\'on de la sem\'antica, luego cuando se instancia con lo del
%  programa ok, pero yo a este nivel no veo lo de usar la font tt. Lo
%  que he usado es la font sf para la direcci\'on de memoria... a ver
%  si os parece bien.}

As an example,  
we show the \KernelC{} rule for assigning a value $V$ of type $T$ to the  variable $X$.
This rule uses three cells: \kcompcell{}, \env{}, and \heap{}. The \env{} cell is a mapping of variable names to their memory positions, whereas the \heap{} cell binds the active memory positions to the actual values. Meanwhile, the \kcompcell{} cell represents a stack of computations waiting to be run, with the left-most (\ie\ top) element of the stack being the next computation to be undertaken. 
%uses two cells, \kcompcell{} and \env{}. 
%The \env{} cell is 
%a mapping of variables to their values, whereas the \kcompcell{} cell represents a stack 
%of computations waiting to be 
%run, with the left-most (\ie\ top) element of the stack being the next computation to be undertaken. 
\[
\setlength{\arraycolsep}{0pt}
\begin{array}{rccclcl}
{\color{white!50!black}\langle}\;&X=\operatorname{tv}(T,V)&\;\ellipses{}{\color{white!50!black}\rangle_{\sf k}}{\color{white!50!black}\langle}\ellipses{}\;X\mapsto{}&\symb{X}&\;\ellipses{}{\color{white!50!black}\rangle_{\sf env}}{\color{white!50!black}\langle}\;\ellipses{}\;\symb{X}\mapsto{}&\_&\;\ellipses{}{\color{white!50!black}\rangle_{\sf heap}}\\[0.4ex]
\cline{2-2}\cline{6-6}\\[-1.6ex]
&\operatorname{tv}(T,V)& & & &\operatorname{tv}(T,V)&
\end{array}
\]

This rule
states that, if the next pending computation  (which may be a part of the evaluation of a bigger expression) consists of an assignment {$X = \operatorname{tv}(T,V)$}, then 
we look for $X$ in the environment ($X\mapsto{}\_)$ %\operatorname{X}$) 
and we update the associated mapping %($\operatorname{X}\mapsto{}\_$) 
in the memory with the new 
value $V$ of type $T$ ($\operatorname{tv}(T,V)$). The 
value {$\operatorname{tv}(T,V)$} is kept at the top of the 
stack (it might be used 
in the evaluation of the bigger expression). 
The rest of the cell's content in the rule does not undergo any modification (this is represented by the $\ellipses$ card).
This example rule reveals a useful feature of \K{}:
\guillemotleft{}rules only need to mention the minimum part of the
configuration that is relevant for their operation\guillemotright{}.
That is, only the cells read or changed by the rule have to be
specified, and, within a cell, it is possible to omit parts of it by simply writing ``$\ellipses{}$''.
For example, the rule above emphasizes the interest 
in: the instruction \mbox{$X = \operatorname{tv}(T,V)$} only at the
beginning of the \kcompcell{} cell, and the mapping from variable $X$ to
its memory pointer {\sf X} at any position in the \env{} cell.
Except for the  subterms that are explicitly identified, upon variable assignment everything is kept unchanged.

%
%A useful feature of \K\ rules is that they are contextual: they
% mention a configuration context in which they apply, together with
% local changes they make to that context. Hence, the developer
% typically only mentions the absolutely necessary information in their
% rules; the remaining details are automatically filled in. 
% 
% For instance,  
The (desugared)  \K\ rule for  \KernelC{} variable assignment is

\begin{center}
\begin{verbatim}
rule   <k> X = tv(T,V) => tv(T,V) ...</k>
       <env>... X |-> pointer(X) ...</env>
       <heap>... pointer(X) |-> (_ => tv(T,V)) ...</heap>
\end{verbatim}
\end{center}

%\maria{Me parece que no est\'an en perfecta correspondencia (AFINAR?).}

%\daniel{Ahora mismo, la regla textual está acorde a la implementación;
%  faltaría sincronizar la regla "pretty print" añadiendo la celda heap
%  y cambiando la posición del término reescrito. Se me ocurre también,
%  para reducir redundancias, que en vez de utilizar la notación tipada
%  tv(T,V) se use directamente V a modo de abstracción.}

% \alicia{creo que lo he arreglado. Al final para pointer(X) he usado
%   la notaci\'on nuestra de usar el sans serif font}

\noindent where the underscore stands for an
 anonymous variable.
 The ellipses are also part of the desugared \K{} syntax and are used to replace the
unnecessary parts of the cells. Hence, also in the desugared rule, the developers
 typically only mention the  information that is absolutely necessary in their
 rules. %; the remaining details are automatically filled in. 
\section{Running Example}\label{sec:runningExample}
% % !TEX root = ./PROLE2015-APV.tex
Our inference technique  relies on the classification scheme
developed in \cite{Liskov86} for data abstractions, %in general, 
where a
function (method) may be either a {\it constructor}, a {\it modifier} or an 
{\it observer}.  A constructor returns a new object of the class from
scratch (i.e., without taking the object as an input parameter).  A
modifier alters an existing class instance (i.e., it changes the state
of one or more of the data attributes in the instance).  An observer
inspects the object and returns a value characterizing one or more of
its state attributes.  We do not assume the traditional premise of the
original classification in \cite{Liskov86} that states that observer
functions do not cause side effects on the state.  This is because we
want to apply our technique to any program, which % that
may be written by third-party software producers that may not follow
the observer purity discipline.
%\alicia{Note that, in our example, {\tt remove} is both an observer
%  (it returns a value different from {\tt void}) and a modifier.}

Let us introduce the leading example that we use to describe the
inference methodology developed in this paper: a \KernelC\
implementation of an abstract datatype for representing doubly-linked
lists. Since the whole example includes a total of 13 methods,   due to
space restrictions we have chosen to comment on just one modifier and five
observer methods (of which 2 are both modifiers and observers).

 \begin{figure*}%[td!]
\lstset{
 language=C,
%frame=single,
% rulesepcolor=\color{blue},
 basicstyle=\ttfamily,
 breaklines=true,
% numbers=left,
 tabsize=1,
columns=[l|l]flexible
}

{\footnotesize
\begin{lstlisting}[multicols=2]
#include <stdlib.h>

struct List {
  void* data;
  struct List* next;
  struct List* prev;
};

struct List* append(struct List* list, void* d) {
  struct List* new_node;
  struct List* final;
  
  new_node = (struct List*) malloc(sizeof(struct List));
  new_node->data = d;
  new_node->next = NULL;

  if (list != NULL) {
     final = list;
     if (final != NULL) {
        while (final->next != NULL)
          final = final->next;
     }
     final->next = new_node;
     new_node->prev = final;

     return list;
  }
  else {
     new_node->prev = NULL;
     list = new_node;
     return list;
  }
}

int length(struct List* list) {
  int len;
  
  len = 0;
  while (list != NULL) {
    len = len + 1;
    list = list->next;
  }
  return len;
}

struct List* reverse(struct List* list) {
  struct List* final;
  
  final = NULL;
  while (list != NULL) {
    final = list;
    list = final->next;
    final->next = final->prev;
    final->prev = list;
  }
  return final;
}

void* head(struct List* list) {
  if (list != NULL) {
      while (list->prev != NULL)
          list = list->prev;
  }
  return list->data;
}

struct List* last(struct List* list) {
  struct List* reversed;
  
  reversed = reverse(list);
  return head(reversed);
}

int find(struct List* list, void*  d) {
  int found;
  
  found = 0;
  while (list != NULL && !(found)) {
      if (list->data == d)
        found = 1;
      else
        list = list->next;
  }
  return found;
}

struct List* init(struct List* list) {
  struct List* aux;

  if (list != NULL) {
     if (list->next != NULL) {
      aux = list->next;
      while (aux->next->next != NULL)
          aux = aux->next;
      aux->next = NULL;
     }
     else 
      list = NULL;
  return list;
}
\end{lstlisting}
%struct List* last (struct List *list) {
%  if (list != NULL) {
%      while (list->next != NULL)
%          list = list->next;
%  }
%  
%  return list;
%}
%
%struct List* remove (struct List *list, void* data) {
%  struct List *tmp;
%  int finished;
%  finished = 0;
%
%  tmp = list;
%  while (tmp && !finished) {
%    if (tmp->data != data)
%       tmp = tmp->next;
%    else {
%       if (tmp->prev)
%          tmp->prev->next = tmp->next;
%       if (tmp->next)
%          tmp->next->prev = tmp->prev;
%       if (list == tmp)
%          list = list->next;
%       free(tmp);
%       finished = 1;
%    }
%  }
%  return list;
%}
}
\medskip
\caption{\KernelC{} implementation of a doubly-linked list.}
\label{DoubleLL}

\end{figure*}

\begin{example}\label{ex:program}
In the \KernelC\ program of Figure~\ref{DoubleLL}, we represent a
doubly-linked list as a data structure ({\tt struct List}) that
contains some content (field {\tt data}), a pointer to the previous
element in the list (field {\tt prev}), and another pointer to the
succesive element in the list (field {\tt next}). %, and a value of any type as the content of the current node (field {\tt data}).

A call {\tt append(list,d)} to the {\tt append} function proceeds as follows: first, a new node {\tt new\_node} is allocated % (represented as a list of one element) 
in memory; it is filled with the value {\tt d} and its {\tt
  next} pointer is initialized to {\tt NULL} since it will become % can not be other than 
the last item in the list.
  %surely will be the last node of the list. % when the procedure ends its execution. 
%
Next, the function checks that the provided list {\tt list} is not {\tt NULL}, in which case it %, if not, 
binds the  {\tt next} pointer of the final element of the list %(resp.\  {\tt prev}) pointer of {\tt
%  list} (resp.\ {\tt new\_node}) 
to    the newly created node, and the {\tt prev} pointer of the new
node to the  final node of {\tt list}, then returns the pointer to the
whole resulting {\tt list}.
 Otherwise, when the input list {\tt list} is null, then the {\tt prev} pointer of {\tt new\_node} is initialized to {\tt NULL} and the resulting full-fledged list that consists of one single element is simply returned.

% \maria{Lo dejo asi provisionalmente para la revisi\'on de Suzanne pero hay algo muy raro
% en la funcion last: si devuelve un puntero a la lista invertida no est\'a devolviendo el \'ultimo elemento, no?
% (estar\'a devolviendo la lista invertida..). Por otra parte, conf\'{\i}o que Dani se las ingeniar\'a para implementar la funci\'on {\sf init} para que devuelva toda la lista menos el ultimo elemento
% sin tener que definir una funci\'on auxiliar...}

The observer function {\tt length}  traverses the list by visiting every node %in the list
in order to count the number of   elements in the list. {%\color{red}
The observer
function {\tt head} returns the data field of the first node of the
list; {\tt last} delivers the data field of the
last node of the list,} which is done by first invoking {\tt reverse(list)} to compute a mirrored version of the parameter {\tt list} and
  then accessing the {\tt data} field of its first node. 
The function {\tt init(list)} returns the same list after removing the  last item of the list.
Finally, the observer {\tt find} looks for the provided  {\tt d} value in the list, and  returns   {\tt 1} (which stands for {\em true}) if the {\tt d} value is found;
%that represents a successful search; 
otherwise, the value {\tt 0} (which stands for {\em false}) is returned. 

\end{example}

From the program code of Example \ref{ex:program}, for each modifier function $m$, we aim to synthesize  an axiomatic specification % 
that consists of a set of 
  implication formulas  %of the form 
  $t_1\Rightarrow t_2$, 
where $t_1$ and $t_2$ are conjunctions of equations of the form $l=r$.
The left-hand side  $l$ of each equation can be either

\begin{itemize}
\item a  call to an observer function 
and then $r$ represents the return value of that call;
\item 
the %label 
 keyword {\tt ret}, and then $r$ represents the   value  returned by the modifier function $m$ being observed.
\end{itemize}

Informally, the statements 
on the left-hand and right-hand sides of the symbol $\Rightarrow$ 
are respectively satisfied before and after the execution of a function call to $m$.
We adopt the standard primed notation for representing variable values after the execution. 
%%% SPACE LIMIT
% For instance, given a variable {\tt s} that stands for the value 
%  of the parameter $s$ 
% before the function is executed, 
% the primed version  {\tt s}'  
% stands for the value after the execution.
%%%%

%\maria{>D\'onde aparecen las primas en la especificaci\'on ejemplo inferida?}
%
\begin{example}\label{ex:specification}
Consider again the program of Example \ref{ex:program}.
The specification for the (modifier) function %{\tt append(list,data)}
{\tt append}   that inserts an element {\tt d} at the end of the list {\tt list}  is shown in Figure \ref{fig:spec}. 
\begin{figure}%[!htd]
%\maria{He metido unos pares de par\'entesis para que se lea mejor.}
\begin{align*}
\begin{minipage}{.40\linewidth}
\small
\[\left(
\begin{array}{l}
\mathtt{length(list)}=\mathtt{0}\; \wedge\\
\mathtt{reverse(list)}=\mathtt{NULL}\; \wedge\\
\mathtt{find(list,d)}=\mathtt{0}\; \wedge\\
\mathtt{init(list)}=\mathtt{NULL}\; \wedge\\
\mathtt{last(list)}=\mathtt{NULL}
%\mathtt{remove}=\mathtt{NULL}
\end{array}
\right)\]
\end{minipage}
& \Rightarrow 
\begin{minipage}{.40\linewidth}
\small
\[\left(
\begin{array}{l}
\mathtt{length(list')}=1\; \wedge\\
\mathtt{reverse(list')}=\mathtt{list}\; \wedge\\
\mathtt{find(list',d)}=\mathtt{1}\; \wedge\\
%\mathtt{head(list')}=\mathtt{list'}\; \wedge\\
\mathtt{init(list')}=\mathtt{NULL}\;\wedge\\ %\mathtt{list}\; \wedge\\
\mathtt{last(list')}=\mathtt{d}\; \wedge\\
\mathtt{ret}=\mathtt{list'}
\end{array}
\right)\]
\end{minipage}
\\%[-2ex]
\begin{minipage}{.40\linewidth}
\small
\[\left(
\begin{array}{l}
\mathtt{length(list)}=x\; \wedge\\
\mathtt{length(list)}>0
%\mathtt{head(list)}=\mathtt{list}
%\mathtt{init(list)}=\mathtt{list}
%\mathtt{reverse}=\mathtt{list}%\; \wedge\\
%\mathtt{remove}=\mathtt{list}
\end{array}
\right)\]
\end{minipage}
& \Rightarrow 
\begin{minipage}{.40\linewidth}
\small
\[\left(
\begin{array}{l}
\mathtt{length(list')}=x+1\; \wedge\\
%\mathtt{reverse}=\mathtt{next}\; \wedge\\
%\mathtt{remove}=\mathtt{list}\; \wedge\\
\mathtt{find(list',d)}=\mathtt{1}\; \wedge\\
%\mathtt{head(list')}=\mathtt{list'}\; \wedge\\
%\mathtt{init(list')}=\mathtt{list}\; \wedge\\
\mathtt{last(list')}=\mathtt{d}\; \wedge\\
\mathtt{ret}=\mathtt{list'}
\end{array}
\right)\]
\end{minipage}
%\begin{minipage}{.25\linewidth}
%\[\left(
%\begin{array}{l}
%\mathtt{length}=1\; \wedge\\
%\mathtt{reverse}=\mathtt{list}%\; \wedge\\
%%\mathtt{remove}=\mathtt{list}
%\end{array}
%\right)\]
%\end{minipage}
%& \Rightarrow 
%\begin{minipage}{.25\linewidth}
%\[\left(
%\begin{array}{l}
%\mathtt{length}=2\; \wedge\\
%\mathtt{reverse}=\mathtt{next}\; \wedge\\
%%\mathtt{remove}=\mathtt{list}\; \wedge\\
%\mathtt{find}=\mathtt{1}\; \wedge\\
%\mathtt{ret}=\mathtt{list}
%\end{array}
%\right)\]
%\end{minipage}
%\\
%\begin{minipage}{.25\linewidth}
%\[\left(
%\begin{array}{l}
%\mathtt{length}=2\; \wedge\\
%\mathtt{reverse}=\mathtt{next}%\; \wedge\\
%%\mathtt{remove}=\mathtt{list}
%\end{array}
%\right)\]
%\end{minipage}
%& \Rightarrow 
%\begin{minipage}{.25\linewidth}
%\[\left(
%\begin{array}{l}
%\mathtt{length}=3\; \wedge\\
%\mathtt{find}=\mathtt{1}\; \wedge\\
%\mathtt{ret}=\mathtt{list}
%\end{array}
%\right)\]
%\end{minipage}
\end{align*}
\vspace{-3ex}
\caption{Expected specification for the {\tt append(list,d)} function  call.}
\label{fig:spec}
\end{figure}
The specification consists of two implications stating the
conditions that are satisfied before and after the execution of a symbolic
%considered
 function call {\tt append(list,d)}.
 %, using the primed notation to distinguish between both states of the modified object.  
 The first formula can be read as
follows: {%\color{red}
if, before executing {\tt append(list,d)},} the result of running {\tt length(list)} is equal to
  0, a call to {\tt find(list,d)} returns 0 (since no value can be found in an empty list) and the results of executing {\tt reverse(list)}, {\tt init(list)}, and {\tt last(list)} are all {\tt NULL}
  % in every case (which confirms that 
  (i.e.,\ the list is empty), 
  then, after executing {\tt append(list,d)}, the length of the augmented list is 1, 
  {%\color{red}
the reversed list coincides with the list itself, 
  % init segment of {\tt list'} is {\tt list},   \maria{(revisar: es muy marciano que el \'ultimo  de una lista sea una lista)} its last element is the one pointed by {\tt list'}, and 
 % reversed list has as its head the node now pointed by {\tt list} and 
 the value {\tt d} can now be found in the list, the init segment
 of the list is {\tt NULL}, the last element is the inserted value
 and the call returns the pointer to the (augmented) list. %non-null {\tt list'}, which means a node was effectively inserted into the empty list.
}
%\maria{El axioma general  no es completo, no? Le falta afirmar que el elemento insertado ser\'a el \'ultimo de la lista,
%es decir, la cabeza de la lista invertida. Creo que nos convendr\'{\i}a m\'as un observador {\tt head}}
The second formula represents the general case: given a  {\tt list}
with an arbitrary size $x$, the call {\tt append(list,d)}
 causes the length to be increased by 1, {%\color{red} 
the inserted value is  found
 in the list, in particular it is returned by the {\tt last} observer, and % {\tt list'}
% , its initial segment is equal to {\tt list}, and
%head node of 
the (augmented) list %{\tt list'} 
is returned. }% as a result.
% \maria{(ESTO   RECHINA UN MONT\'ON:) Moreover, the last node of the list has changed and corresponds to a fresh pointer identified with an arbitrary number.}
Since the {\tt append} function does not restrict the insertion to the cases in which the {\tt d} value is still not inside the list, we cannot assume {\tt find} to return 0 before running the modifier function {\tt append}.

  %and {\tt remove(list,data)} returns the same pointer {\tt list} (which means
%that the resulting list starts by the same element, pointed-to by {\tt list}),%  it does not eliminate the only node inside the list and, as a
% consequence, the value {\tt data} is still not inserted)
%then, after executing the method, the length is increased to 2,
%and the reversed list (obtained from the call {\tt reverse(list)})
%has as its head the node {\tt list.next}.  Since the new node is
%inserted at the end, the value returned by the call is the same
%pointer {\tt list}.  

%Even though the observers {\tt isnull} and {\tt isfull} behave as boolean functions 
%(predicates) in this example, we prefer not to
%write them in sugared relational form (\ie{}   {\tt isfull(s)} instead of 
%{\tt isfull(s)=1} )
% since a specific  datatype for Boolean numbers does not exist in \C{}. 
%Hence, even when we can detect  that the observer function only  returns two scalar values, say  {\tt 0} and {\tt 1} as in the example, 
%we cannot give it the semantics of a logical predicate.

% \maria{Falta implementar el c\'odigo del \tt{init(5)}. Por otro lado, no entiendo por qu\'e sale el \tt{FreeInt(5)} en el axioma 2 en el caso del \tt{last} (?)}

\end{example}
Note that any implication formula in the specification may contain multiple facts (in the pre- or post-condition) that refer to function calls that are assumed to be run independently under the same initial %state 
conditions. This avoids making any assumptions about  function purity or side-effects.

%In the following section  we describe our technique for inferring specifications by using \K's symbolic execution engine.
 
%the   \MLlong{} verifier \MatchC{}.
% \MatchC{}  works in a forward manner by symbolically executing an \ML\ pattern 
% that is provided as the program precondition, and non-deterministically obtaining a set of final patterns that are then used to discharge the postcondition.
%This is an instance of a general strategy to calculate the strongest postcondition of a predicate transformation semantics as explained in~\cite{FW-HOARE2010}.
%However, \MatchC\ is incomplete for the purpose of  general symbolic execution in the sense that its symbolic  machinery does not support incremental  assumptions regarding the initial structure of the 
%program memory; 
% it can only assume the structure 
%that is implicitly imposed by the initial pattern.
%For the inference purposes of this paper, we cannot assume any {\em ex-ante} condition for the initial 
%program state; on the contrary, we need to incrementally collect all the assumptions that allow each symbolic execution  path  to be successfully  executed. 
%In the following section, we explain how we  extended \MatchC{} 
%to support collecting assumptions on-the-fly within 
%the symbolic configurations as needed.

%% %----------------------------------SECTION-----------------------------------%

\section{Symbolic Execution in the \K Framework}\label{sec:SEinK}
% !TEX root = ./PROLE2015-APV.tex
%\label{sec:symbExec}

Symbolic execution consists of executing programs with symbolic values instead of concrete values.
%Symbolic execution typically 
It proceeds like standard execution except
that, when a function or routine is called, symbolic values are
assigned to the actual parameters of the call and computed values become symbolic
expressions that record all operations being applied.
When symbolic execution reaches a conditional control flow statement,
every possible execution path from this execution point must be explored.
In order to keep track of the explored execution paths, symbolic
execution also records the assumed (symbolic) conditions on the
program inputs that determine each execution path in the so-called
\emph{path conditions} (one per possible branch), which are empty at
the beginning of the execution.
A path condition consists of the set of constraints that the arguments
of a given function must satisfy in order for a concrete execution of
the function to follow the considered path.
Without loss of generality, we assume that the symbolically executed
functions access no global variables; they could be easily modeled by
passing them as additional function arguments.

\begin{example}\label{ex:3}
Consider again the {\tt append} 
  function  of Example~\ref{ex:program}.
Assume that the input values for the actual parameters {\tt list} and {\tt d} are the symbolic pointer \symb{list} and the symbolic value \symb{?d}, respectively. 
Then, when the symbolic execution reaches the first {\tt if} statement in the code, 
it explores the two paths arising from considering both the satisfaction and non-satisfaction of the guard in the conditional branching statement. 
The 
path condition of the first branch is updated with the constraint $\symb{list}\neq{\tt NULL}$, whereas $\symb{list}={\tt NULL}$ is added to the path condition in the second branch.
\end{example}

To summarize, symbolic execution can be represented as a tree-like structure where each branch corresponds to a possible execution path and has an  associated  
path condition.
The {\em successful} paths are those leading to a final (symbolic) configuration that encloses a satisfiable path constraint
and that typically stores  a (symbolic) computed result.

For the symbolic execution of \KernelC\, programs, we must pay attention to pointer dereference and initialization.
In  \C{}, a structured datatype ({\tt struct}) is an aggregate type that is used to comprise  a nonempty set of sequentially allocated member objects\footnote{An object in \C{} is a region of data storage in the execution environment.},
called fields, each of which has a name and a type.
When a {\tt struct} value is created, \C{} uses the address of its first field to refer to the whole structure.
In order to access a specific field {\tt f} of the 
given structure type,
\C{} computes {\tt f}'s address by adding an offset (the sum of the sizes of each preceding field in the definition) to the address of the whole structure.

In our symbolic setting,   the pointer arithmetics and memory layout
machinery are abstracted by 1) using symbolic variables as addresses, and 2)
mapping each structure object into a single element of the \heap{} cell that groups all object fields (and associated values). % , as if all fields of an object were stored in the same memory location.
  A specific field is then accessed by combining the identifiers of both the structure object and the field name.

%\alicia{NOTA: con esta aproximaci\'on no podemos considerar el aliasing debido a la
%  aritm\'etica de punteros (desde un objeto acceder a otro en una
%  posici\'on de memoria a continuaci\'on). Solo es un comentario, por
%  tenerlo claro, no porque quiera cambiarlo.}

% \alicia{revisar si tt o sf para los punteros first y third (en el
%   texto aparecen como tt pero en la configuraci\'on como sf)}
  
% \daniel{Es sf al tratarse de punteros simbólicos. La duda que tengo yo es si los campos deberían ser tt o sf, aunque votaría por tt puesto que proceden del código y así evitamos abusar de la sans-serif.}

\begin{example}\label{ex:4}
Consider the structure type {\tt List} of Example
\ref{ex:program}. The following configuration records a list variable
\prgvar{l} with: 1) the integer 7 in its {\tt data} field; 2) a
reference (pointer) named \symb{first\_node} as the value of its {\tt
  prev} field; and 3) a reference (pointer) \symb{third\_node} as the value of its {\tt next} field:
%whose with value 7 in its {\tt capacity} field, value 1 in its {\tt size} field and a reference to a pointer named {\tt first\_node} as value of its {\tt elems} field will be recorded in the configuration as follows:
\[
\kallLarge{cfg}{
\ldots{}
\kall{env}{
	\mapsTo{\prgvar{l}}{\symb{l}}
}
\kmiddle{heap}{
	\mapsTo{\symb{l}}{(\mapsTo{\prgvar{%\color{red}
data}}{\operatorname{tv}(int,7)}, \mapsTo{\prgvar{%\color{red}
prev}}{\symb{first\_node}}, \mapsTo{\prgvar{%\color{red}
next}}{\symb{third\_node}})}
}
\ldots{}
}
\]
In order to access a field of the list \prgvar{l}  (e.g., its
{\tt data} field), the corresponding index is computed by juxtaposing
the identifier of the {\tt data} field to the pointer \symb{l}, thus mimicking how the concrete access
would be done in \C{} (\ie{}
{\tt l->data}). 
% \alicia{yo estoy lo quitar\'{\i}a \\ In this way, the actual
%   distribution of stored values in different memory positions is
%   abstracted to both the user and the system, which improves
%   comprehension and simplifies both the output and the computation.
% }

% \daniel{Concuerdo. Aunque el mensaje que se intenta transmitir sí que creo que sería interesante para justificar por qué hemos tomado esta opción en lugar de la aritmética de punteros tradicional, aunque sea más conocida, la forma en que lo expresé puede dar lugar a confusión. Además, ya hemos comentado antes que es una abstracción, también sería repetir cosas.}
%Consider the second {\tt if} statement of the {\tt add} function given in  Example~\ref{ex:program}.
%The evaluation of the guard of the conditional statement requires accessing both
%{\tt p->size} and {\tt p->capacity}. Since {\tt capacity} is the first field in the {\tt struct set} definition, 
%its location coincides with the (base) address {\tt p}.
%However, 
%in order to access {\tt p->size}, its address 
%must be computed
%by adding an offset\footnote{We assume that the memory is indexed by words and that a value of type {\tt int} has the size of a word.} 
%of {\tt 1\/} to the (base) address {\tt p} (\ie{} if we assume that the symbolic address held by the variable {\tt p} is \symb{p}, then the computed address for the field {\tt size} is 
%\symb{p}$+1$.)
\end{example}

Another critical point is 
the \emph{undefinedness} 
problem that occurs in \C{}  programs when accessing uninitialized memory addresses.
The \KernelC{} semantics that we use preserves the concrete \emph{well-definedness} behavior of pointer-based program functions of \C\ while still detecting the \emph{undefinedness} cases in a way similar to the \C\ operational semantics of~\cite{ellison-rosu-2012-popl}.
However, in our   
inference setting, we have no a priori information regarding the memory (specifically, information about the (un)initialized memory addresses).
Therefore, when symbolic execution accesses (potentially uninitialized) memory positions, two cases must be considered:
the case in which the memory is actually initialized and stores an
object, and the case in which it stores a null pointer. %\alicia{esto de
%  abajo es
%  as\'{\i}? porque el tercer axioma se corresponde con s\'{\i}
%  considerarlos cuando se accede a esos objetos, en la guarda se
%  considera que sea null o que no lo sea... no s\'e si he entendido
%  mal la explicaci\'on\\
%Ahora que he seguido leyendo creo que lo tengo claro, pero creo que
%confunde, que mejor quitarlo (en papel me hab\'{\i}a apuntado eliminar
%la frase que hab\'{\i}a originalmente en vez de intentar explicarlo mejor}
In contrast to the approach described in \cite{AFV13}, we do not consider cases where the
  pointer is %uninitialized \maria{or?} 
  undefined (\ie{} when the execution is halted due to forbidden pointer access). This avoids accumulating too
  many solutions with undefined behavior, which could cause an
  explosion of axioms for programs that access new objects frequently, resulting in huge and redundant output specifications.
For the 
case when the memory positions are actually initialized with non-null objects, %and execution should proceed, 
a strategy to reconstruct the original object in memory is needed. We adapt the lazy initialization of objects of~\cite{GenSymbExec2003} to our setting:
when a symbolic address (or address expression) is accessed for the first time, \SE{} initializes the memory object 
that is located at the given  address with a new symbolic value.
This means that the mapping in the \heap{} cell is updated by assigning a new free variable to the symbolic address
of the accessed field so that, from that point on, accesses to that field can only succeed. As a result, {\em undefined} computations can only 
%be final states 
occur in the case of syntactic program errors (\ie{}   expressions that are not accepted in the specification of the language).
%the case in which the memory is actually initialized, 
%and the case in which it is not.
%\alicia{revisar \\
%In the second case, the symbolic execution gets stuck, 
%thus identifying \emph{undefined} behavior as in~\cite{ellison-rosu-2012-popl}.}
%\alicia{inluir aliasing? \\
%For the 
%case in which the memory positions are actually initialized and execution should proceed, 
%a strategy to reconstruct the original object in memory is needed. 
%We adapt  to our setting the lazy initialization of fields of~\cite{GenSymbExec2003}:
%when a symbolic address (or address expression) is accessed for the first time, \SE{} initializes the memory object 
%that is located at that address with a new symbolic value.
%This means that the mapping in the \heap{} cell is updated by assigning a new free variable to the symbolic address
%of the accessed field so that from that point on, accesses to that field can only succeed.}
% In contrast, in the case of failure, an {\tt undefined} computation is pushed onto the \kcompcell{} cell, which stops the execution.

% \alicia{TO DO (para pr\'oxima versi\'on) Creo que se entender\'{\i}a mejor la LI si us\'aramos como
%   ejemplo una instrucci\'on que no fuera una expresi\'on booleana, por
% ejemplo el list = list->next del método length. As\'{\i}
% explicar\'{\i}amos que se bifurca incluso en una asignaci\'on.}
\begin{example}(Example {\ref{ex:3} continued})
Before executing the first {\tt if} statement for the first time, assume that the \heap{} cell is empty,
which means that nothing is known about the structure of the \heap{}.
After symbolically executing the guard of the {\tt while} statement (which refers to the {\tt next} field of the structured data   {\tt final}),
by applying the lazy initialization approach,
the \heap{} cell gets updated to: \sePostCondIfTwo{}%
In other words, new symbolic bindings for the actual parameters are added, which represent the assumptions we made over 
the corresponding data structures. More specifically, the accessed
field is initialized with a fresh symbolic pointer \symb{list.next}
whereas the fields that have not been accessed yet (temporarily)
remain  % in an 
undefined, % state, 
in a state that is specified by the symbolic constant %characterized by the term 
$\operatorname{undef}$. 
% \alicia{lo quito de momento porque no me queda claro

% To avoid inconsistencies, an undefined object is created and linked to that new symbolic pointer whose fields will be initialized (if required) in future computation steps.}
\end{example}

In the following section, we describe \K's symbolic execution
machinery and how we adapted it to support discovering program specifications.

\subsection{The symbolic machinery in \K}\label{sec:technicalSE}

Recently, the \K{} framework has been enriched with a tool that
automatically compiles language definitions into symbolic
semantics. In other words, any language that is formally defined in   \K\
%definition 
can (ideally) benefit, without cost, from symbolic execution.
The \K\ symbolic backend automatically {%\color{red}
attaches to the
configuration} a new cell, called \pathcondcell{}, for the conditions
on the input arguments that are accumulated during the symbolic
execution.
Roughly speaking, the mechanism works as follows: whenever a non-deterministic choice is found 
%in the rule set of the specification, 
(\ie{}   the term at the top of the
\kcompcell{} cell can be rewritten by applying different rules), the symbolic engine
considers each path independently, storing 
the assumptions that enable each concrete
execution  path in the \pathcondcell{} cell.
Therefore, the symbolic execution of  programs under the \K{} framework 
results in a set of \emph{patterns} (consisting of the final symbolic
configuration that encloses the corresponding  \pathcondcell{} cell) which we call \emph{final
  patterns}.

%The following example illustrates the \emph{modus operandi} described above. 

\begin{example}\label{ex:kpathcond}
Assume that our \K specification contains these two rules, which represent the possible rewritings of an {\tt if} statement:
\medskip

	%rule if(true) S:Statement else _:Statement => S
	%rule if(false) _:Statement else S:Statement => S
%\begin{verbatim}
%	rule if(true) S else _ => S
%	rule if(false) _ else S => S
%\end{verbatim}
\begin{tabular}{cc}
\begin{minipage}{.4\linewidth}
\[
\setlength{\arraycolsep}{0pt}
\begin{array}{rcccl}
{\color{white!50!black}\langle}\;&\mathrm{if}\;({\tt true}) \; S  \; \mathrm{else} \; \_ &\;\ellipses{}{\color{white!50!black}\rangle_{\sf k}}%{\color{white!50!black}\langle}\ellipses{}\;X\mapsto{}&\_&\;\ellipses{}{\color{white!50!black}\rangle_{\sf env}}
\\[0.4ex]
\cline{2-2}%\cline{4-4}
\\[-1.6ex]
%&\operatorname{tv}(T,V)& &V&
&S&
\end{array}
\]
\end{minipage}
&
\begin{minipage}{.4\linewidth}
\[
\setlength{\arraycolsep}{0pt}
\begin{array}{rcccl}
{\color{white!50!black}\langle}\;&\mathrm{if}\;({\tt false}) \; \_  \; \mathrm{else} \; S &\;\ellipses{}{\color{white!50!black}\rangle_{\sf k}}%{\color{white!50!black}\langle}\ellipses{}\;X\mapsto{}&\_&\;\ellipses{}{\color{white!50!black}\rangle_{\sf env}}
\\[0.4ex]
\cline{2-2}%\cline{4-4}
\\[-1.6ex]
%&\operatorname{tv}(T,V)& &V&
&S&
\end{array}
\]
\end{minipage}
\end{tabular}

Now assume that we are running the following piece of code:
\begin{verbatim}
if (x > y) return 1; else return 0;
\end{verbatim}
with symbolic variables  {\tt x} and {\tt y}, and no initial
restrictions over them (\ie{} the \pathcondcell{} cell is initialized
to {\tt true}).
%
%We don't know if the boolean expression will evaluate to {\tt true} or
%to {\tt false}; thus, we have non-determinism within the rules of the
%specification. 
The compilation of the language with \K's symbolic
backend explores both branches (i.e.,\ the   case when the guard {\tt x > y} is true and the case when the guard evaluates to  false),
%, triggering a bifurcation in the
%execution tree and obtaining 
which respectively lead to the following patterns\footnote{We only write those cells that are relevant for the
  example.}:
\medskip

%\begin{verbatim}
Branch 1:
%\end{verbatim}
$
\kallLarge{\cfg{}}{\footnotesize
 \kall{k}{
   \operatorname{tv}(int,1)
 }
  \ldots{}
 }
%\\
\kallLarge{\pathcondcell}{\footnotesize \symb{?x} > \symb{?y}} % = \operatorname{true}}
$
\medskip

%\begin{verbatim}
Branch 2:
%\end{verbatim}
$
\kallLarge{\cfg{}}{\footnotesize
 \kall{k}{
   \operatorname{tv}(int,0)
 }
  \ldots{}
 }
%\\
\kallLarge{\pathcondcell}{\footnotesize \symb{?x} \leq \symb{?y}} % = \operatorname{false}}
$
\end{example}

%Unfortunately, the versions of \K{} framework that possess this capability do not have a compatible \KernelC{} specification associated. Under these circumstances, we had to adapt a previous installment of the language %specification to the recent symbolic engine and to our requirements.
 
As already mentioned, the exhaustive symbolic execution of all paths cannot always be achieved in practice because an unbounded number of paths can arise in the presence of loops or recursion. We follow the standard approach to avoid the exponential blowup that is inherent in path enumeration by exploring loops up to a specified number of unfoldings.
This ensures that \SE{} ends for all explored paths, thus delivering a finite (partial) represention of the program behavior  \cite{DSilvaKW08}.
Obviously, not all the potential execution paths are feasible, but \K
deals with this automatically and transparently to the user 
by using the theorem prover Z3 \cite{Z3-2008} 
to check the satisfiability of the path condition constraints.
%and to eliminate unfeasible symbolic computations whenever the corresponding path condition is unsatisfiable (pruning the execution tree).
 
{%\color{red}
It is important to note that the symbolic \K{} engine   is not endowed
with the lazy initialization technique. 
As a consequence, any branching in \K's symbolic execution trees is
associated to the evaluation of a guarded instruction (conditional,
while loop, etc.), whereas lazy initialization also adds bifurcations
when mimicking the access to complex
data structures (objects) because all possible scenarios are
considered. %explicitely (for instance in an assignment instruction). % following the lazy
% initialization approach. 
%
In other words, %with lazy initialization bifurcation 
% More specifically, it does not  automatically store  the path conditions associated to  lazy initialization of structures 
% since the constraints may contain % inferred do
branching is not only caused by the evaluation of guards (boolean expressions), but also by other kinds
of expressions (for instance when assigning a value to a data structure). % with symbolic variables in the \kcompcell{} cell (like the $\symb{x} > \symb{y} = {\tt true}$ in the Example \ref{ex:kpathcond}, which was the guard of an {\tt if} statement) but from
% bindings between some pointers and (previously) undefined values in the
% \heap{} cell. %instead.%}

Note that  the   \pathcondcell{} cell (where  
constraints associated to guards % for the \emph{non complex} (i.e.,\ scalar) data
are stored) is not
under our control but is automatically handled by the \K\  
symbolic engine, which ensures   language independence.  For this
reason, we have adopted the solution to
% Furthermore, we do not have any control over the
% newly added \pathcondcell{} cell because it is independent of the
% configuration, and does not exist at compilation time. To solve this,
% our implementation
introduce a new cell, {%\color{red}
called \mempc{},
into the
configuration  that is used to store  those constraints associated to
non-guarded instructions that refer to
complex data structures. }
By abuse, when we refer to the path condition $\phi$ of a pattern, we
implicitly consider that $\phi$ includes the constraints in \mempc{} as well.

In the following section, we formulate our symbolic specification inference algorithm.

% that symbolically executes the program and automatically  extracts 
%and combines the pre-call and post-call patterns in order to infer the 
%pursued logical specifications.
 
% 

% %----------------------------------SECTION-----------------------------------%
%\section{Inference process}
\section{Inferring Specifications Using \K's Symbolic Execution}
\label{sec:inference}

% !TEX root = ./PROLE2015-APV.tex 
Let us introduce the basic notions that we use in our formalization.
Given an input program, let $\methods$ be the set of functions in the
program.  We distinguish the set of observers $\obsM$ and the set of
modifiers $\modM$.  
A function can be considered to be an \emph{observer} if it explicitly
returns a value, whereas any method can be considered to be a
\emph{modifier}.  Thus, the set $\obsM\cap\modM$ is generally non empty. 
For instance,  the function \texttt{reverse}  in Example~\ref{ex:program} is both
an observer and a modifier function.

Given a function $f\in\mathcal{F}$, we represent a call to $f$ with
the list of arguments $\mathit{args}$ by $f(\mathit{args})$.  Then,
$f(\mathit{args})\{\phi\}$ is the \K pattern built by inserting
the call $f(\mathit{args})$ at the top of the \kcompcell{} cell and
initializing the path condition cells %\pathcondcell{} cell
with $\phi$.  
This is helpful to start the execution of $f(\mathit{args})$ under the (possibly non--empty) constraints of $\phi$.
%$f(\mathit{args})\{{\tt true}\}$ stands for the \K\ pattern
%that represents the execution of $f$ with arguments $\mathit{args}$
%under an initial state without constraints. % , which will normally
% correspond to the initial call of the inference process.
We also denote by $\textsc{Se}(f(\mathit{args})\{\phi\})$ the set of
final patterns obtained from the symbolic execution %in \K\ 
of the pattern
$f(\mathit{args})\{\phi\}$ (\ie{} the leaves of
the deployed symbolic execution tree).
% {\color{red} Given a pattern $p$, $\mathit{pathc}(p)$ extracts the
%   content in
%   the \pathcondcell{} cell of the pattern whereas $\mathit{pathd}(p)$
% extracts the content in our auxiliar path condition cell (that
% associated to the complex data structures). Both conditions are
% simplified via Z3.}

Our specification inference methodology is formalized in
Algorithm~\ref{alg:inference}.
\begin{algorithm}%[htd!]                      % enter the algorithm environment
    \algsetup{linenodelimiter=.}
    
    \begin{algorithmic}[1]                    % enter the algorithmic environment     
      \REQUIRE $m\in \mathcal{M}$ of arity $n$; 
\STATE $S = \textsc{Se}(%\extpat{
m(\symb{a}_1,\ldots,\symb{a}_n)\{{\tt true}\}%}
)$
      \STATE $\mathit{axiomSet}$ := $\emptyset$;
      \FORALL {${p}\in S$ with \pathcondcell{} cell $\phi_p$, 
\mempc{}
        cell $\varphi$ and
                               %and 
return value $\mathsf{v}$}%\Se{}(m[\cdot{}])$}
        \STATE $\!\!\!\!eqs_{pre}$ := $\explain{\kall{\cfg{}}{\footnotesize
 \kall{k}{m(\symb{a}_1,\ldots,\symb{a}_n)
 }\kall{heap}{\varphi}
  \ldots{}
 }
\kall{\pathcondcell}{\phi_p}
,[\symb{a}_1,\ldots,\symb{a}_n]}$;
        \STATE $\!\!\!\!eqs_{post}$ := $\explain{{p},[\symb{a}_1,\ldots,\symb{a}_n]}$; 
        \STATE $\!\!\!\!eq_{ret}$ := ($ret=\mathsf{v}$); 
%        \STATE $\!\!\!\!eq_{ret}$ := $ret=\mathsf{k}(p)$; 
        \STATE $\!\!\!\!\mathit{axiomSet}$ := $\mathit{axiomSet}\cup \{eqs_{pre} \Rightarrow (eqs_{post}\cup{}eq_{ret})\}$;
      \ENDFOR
      \STATE spec := $\mathit{simplify}$($\mathit{axiomSet}$) 
      \RETURN spec
    \end{algorithmic}
    \caption{Specification Inference.} % give
    \label{alg:inference}%the algorithm a caption
\end{algorithm}
First, the \emph{modifier} method of interest $m$ is symbolically
executed with fresh symbolic variables $\symb{a}_1,\ldots,\symb{a}_n$
as arguments and empty constraint {\tt true}, and the set $S$  of final
patterns is retrieved from the leaves of the symbolic execution tree.
%
%Before moving on to the next step, 
For each pattern in $S$, the corresponding path condition is %preprocessed and
simplified (by calling the automated theorem prover Z3) to avoid
redundancies and simplify the analysis.
Then, we proceed to compute an axiom  for each pattern $p$ %(with
                                %simplified path condition $\phi$) 
in $S$ that 
explains (by using the observers) the properties  that hold in the state before and after the execution of the
method. This is done by means of the function $\explain{q,as}$, where $q$ is a pattern and $as$ is a list of symbolic variables, given
in Algorithm~\ref{alg:explanations}. The explanation for  initial states
whose symbolic execution end in $p$ (line 4)  must ensure that  the input data   comply with the
conditions that make the path to $p$ feasible, which is achieved by  % setting
imposing  the conditions given by $\phi$ to  the input pattern to be explained (i.e.\ by feeding its \heap{} cell with $\varphi$ before invoking the routine $\mathit{explain}$).
%that were accumulated during the symbolic
%execution of $m(\symb{a}_1,\ldots,\symb{a}_n)\{{\tt true}\}$. % by extracting and
% processing the pre-call and post-call patterns of each
% $\extpat{p}\in S$, \ie{} the \K patterns representing the state
Then, we proceed to explain (also with the observers) the properties
of the considered final state (the final pattern $p$) by  invoking
$\explain{{p},[\symb{a}_1,\ldots,\symb{a}_n]}$. Finally, the return
value {\sf v}   is retrieved from the \kcompcell{} cell of $p$, and the axiom $ret=\mathsf{v}$ is added to  the specification inferred.
% ($\mathsf{k}(p)$). 
%
This value could be either   undefined   or a single typed
value that represents the return from the function $m$ under the
conditions given by $\phi$.
% in the \pathcondcell{} cell of the pattern.

% before and after the execution of
% $m(\symb{a}_1,\ldots,\symb{a}_n)[\extpat{p}]$, a set of axioms is
% obtained that defines the behaviour of the program.  This is done by
% means of the function $\explains{p,as}$ given in
% Algorithm~\ref{alg:explanations}. 
The computed axioms are
implications of the form $l_i \Rightarrow r_i$, where $l_i$ is a
conjunction of preconditions and $r_i$ is a conjunction of
postconditions. Note that  a conjunction  of equations is represented  as an equation set in Algorithm \ref{alg:explanations}. The function $\mathit{simplify}$ implements a
post-processing which consists of: (1) disjoining the preconditions $l_i$
that have the same postcondition $r_i$ and simplifying the resulting
precondition; and (2) conjoining the postconditions $r_i$ that share the
same precondition and simplifying the resulting postcondition.  % Finally,
% the returned value from each post-call pattern is directly obtained
% from its \kcompcell{} cell, bound to the label $ret$, and the
% resulting equation $eq_{ret}$ is added to the postcondition set
% $r_i$.
% It could be either an undefined computation or a single typed
% value representing the return from the function $m$ under the
% conditions specified in the \pathcondcell{} cell of the pattern.

Let us illustrate the application of the inference algorithm with the following example.

\begin{example}\label{ex:SEmodifier}
Let us compute a %Assume that we want to infer a 
specification for the {\tt append} \emph{modifier} function of Example~\ref{ex:program}.
Following the algorithm, we first compute $\Se{}(\texttt{\tt
  append(list,d)}[{\tt true]})$ with {\tt list} and {\tt d} % \footnote{In this example, we use {\tt d} instead of {\tt data} to avoid confusion with the field {\tt data} in objects of the structured type {\tt List}.}
(free) symbolic variables.
Since there are no constraints in the initial symbolic configuration, the execution covers all possible initial concrete configurations.
For simplicity, we set the number of loop unrollings to one; as a consequence, the symbolic execution computes 
three final patterns. %\footnote{For simplicity, we set the number of loop unrollings to one.}. 
The following pattern $e$ represents the final state for the path where the body of the {\tt while} statement never gets executed (0 iterations):
\[
\kallLarge{cfg}{
 \kall{k}{
   {\scriptsize \smalloperator{tv}\text{(struct \; List*, \symb{list})}}
 }
 \\
  \kall{env}{
    {\scriptsize \mapsTo{\prgvar{list}}{\symb{list}}},
    {\scriptsize \mapsTo{\prgvar{d}}{\symb{d}}},
    {\scriptsize \mapsTo{\prgvar{new\_node}}{\symb{new\_node}}},
    {\scriptsize \mapsTo{\prgvar{final}}{\symb{list}}}
  }\\%[-.5ex]
  \kallLarge{heap}{
    {\scriptsize \mapsTo{\symb{list}}{(\mapsTo{\prgvar{data}}{\smalloperator{undef}},\mapsTo{\prgvar{prev}}{\smalloperator{undef}}, \mapsTo{\prgvar{next}}{\symb{new\_node}})}}
    \\
    {\scriptsize \mapsTo{\symb{new\_node}}{(\mapsTo{\prgvar{data}}{\symb{d}},\mapsTo{\prgvar{prev}}{\symb{list}}, \mapsTo{\prgvar{next}}{\symb{NULL}})}}
    \\
    {\scriptsize \mapsTo{\symb{d}}{\smalloperator{tv}\text{(void, \symb{?d})}}}
  }
 }
\\
\kallLarge{\mempc{}}{
{\scriptsize\symb{list} \neq \symb{NULL} \; \wedge}
\\
{\scriptsize \symb{list}\Rightarrow\prgvar{next} = \symb{NULL}}}
\]

The execution of this path returns the pointer to the resulting list: the returned pointer is represented by the typed value $\operatorname{tv}(struct \; List*, \symb{list})$ in the \kcompcell{} cell.
The field ${\sf list}\Rightarrow{\tt next}$ is accessed only after checking that
\symb{list} is not null: it has been assumed %that 
{\tt list != NULL} at
the first conditional expression, thus the constraint $\symb{list} \neq
\symb{NULL}$ has been added to the path condition, whereas {\tt
  final->next != NULL} (the guard of the {\tt while} loop) is assumed
false, thus the constraint $\symb{list} \Rightarrow {\tt next} = \symb{NULL}$ has been
gathered. Note that, although the variable accessed in the code is {\tt final}, the generated path constraint refers to the pointer \symb{list} since both {\tt final} and {\tt list} are bound to the same memory address in the environment.
\end{example}

Let us now describe 
Algorithm~\ref{alg:explanations} which defines the function $\explain{q,as}$.
Given a \K pattern $q$ and a list of symbolic variables $as$, this
function describes $q$ as a set of equations that are obtained by executing the observer functions in the state.  Each
equation  relates the call to an observer function  (or  
\emph{built-in} function) with the (symbolic) value that
the call returns. In the algorithm, $\mathit{As}\sqsubseteq
\mathit{as}$  means that the list of elements $\mathit{As}$ is a
permutation of some (or all) elements in $\mathit{as}$.

\begin{algorithm}%[htd!]                      % enter the algorithm environment
    \algsetup{linenodelimiter=.}
    
    \begin{algorithmic}[1]                    % enter the algorithmic environment
      \REQUIRE $q$ : the pattern to be explained (with path condition
      $\phi$) % in the \heap{}  cell)
      \REQUIRE       $\mathit{as}$ : a list of symbolic variables
      \STATE $\mathcal{C}$: the universe of observer calls;
      \STATE $\mathit{eqSet}$ := $\emptyset$;
      \FORALL{$o(\mathit{As})\in\mathcal{C}$ with $\mathit{As}\sqsubseteq \mathit{as}$}
%        \STATE $S = \Se{}(o(\mathit{As})\{\phi(q)\})$
         \STATE $S = \Se{}(o(\mathit{As})\{\phi\})$
       \IF{ $\nexists{}~{q_1},{q_2}\in S$ 
              s.t. $q_1  \mbox{ and } q_2 \mbox{ contain a different return value } \mathsf{k} \mbox{ in their } \kcompcell{} \mbox{ cell }$  
       %s.t. $\mathsf{k}({q_1}) \neq{} \mathsf{k}({q_2})$ 
       }
%s.t. $\mathsf{k}({q_1}) \neq{} \mathsf{k}({q_2})$ }
%          \STATE $\mathit{eqSet}$ := $\mathit{eqSet} \cup (o(\mathit{As})=\mathsf{k}({q_1}))$
          \STATE $\mathit{eqSet}$ := $\mathit{eqSet} \cup (o(\mathit{As})=\mathsf{k})$
        \ENDIF
      \ENDFOR
      \RETURN $\mathit{eqSet}$
    \end{algorithmic}
    \caption{Computing explanations: $\explain{q,\mathit{as}}$}
    \label{alg:explanations}%the algorithm a caption
\end{algorithm}

Roughly speaking, given a pattern $q$, $\explain{q,\mathit{as}}$ first generates the universe of observer function calls $\mathcal{C}$, which consists of all the function calls $o(\mathit{As})$ that satisfy that: 
\begin{itemize}
\item $o$ belongs to $\mathcal{O}$ or to the set of (predefined) built-in functions,
\item the argument list $\mathit{As} \sqsubseteq \mathit{as}$ %in the call $o(\mathit{As})$
% is a suitable selection of variables from
%the symbolic variable list $\mathit{as}$ that is received as argument, respecting
respects the type and arity of $o$. 
\end{itemize}

Then, for each call $o(\mathit{As}) \in \mathcal{C}$, Algorithm
\ref{alg:explanations} checks whether all the final symbolic
configurations (leaves) resulting from the symbolic execution of
$o(\mathit{As})$,  under the constraints given by %conditions 
$\phi$,  % allocated in the $\heap{}$ cell of $q$
 have 
the same return value. 
When the call satisfies this requirement, 
an equation is generated (line 6 in Algorithm~\ref{alg:explanations}). Otherwise, the observation is inconclusive and no explanation is delivered in terms 
of the executed observer function.
% The intuition of this step is that, if we symbolically execute the observer at a given initial state and,  for all its execution branches, we get  the same value, then the observer 
% together with the   return value (partially) characterize the considered  state.
The algorithm finally returns 
the set of all the  explanatory equations inferred. 

\begin{example}[Example~\ref{ex:SEmodifier} continued]
Let us show how we compute the explanation for the final state of
% represented on the 
pattern $p$ in Example~\ref{ex:SEmodifier}. % given the  symbolic variables considered in the example.
Given the observer functions {\tt length}, {\tt reverse}, {\tt head}, {\tt last}, {\tt find}, and {\tt init}, and the symbolic variables {\tt list} and {\tt d}, the universe of observer calls is {\tt length(list)}, {\tt reverse(list)}, {\tt head(list)}, {\tt last(list)}, {\tt find(list,d)}, and {\tt init(list)}.
Let us consider the case for the observer call {\tt length(list)} in detail.

When we symbolically execute {\tt length(list)} on the %considered 
pattern $p$, we obtain a single final pattern:
\[
\kallLarge{cfg}{
 \kall{k}{
   {\scriptsize \smalloperator{tv}\text{(int, \em{2})}}
 }
 \\
  \kall{env}{
    {\scriptsize \mapsTo{\prgvar{list}}{\symb{list}}},
    {\scriptsize \mapsTo{\prgvar{length}}{\symb{length}}}
  }\\%[-.5ex]
  \kallLarge{heap}{
    {\scriptsize \mapsTo{\symb{list}}{(\mapsTo{\prgvar{data}}{\text{\smalloperator{undef}}},\mapsTo{\prgvar{prev}}{\smalloperator{undef}}, \mapsTo{\prgvar{next}}{\symb{new\_node}})}}
    \\
    {\scriptsize \mapsTo{\symb{new\_node}}{(\mapsTo{\prgvar{data}}{\symb{d}},\mapsTo{\prgvar{prev}}{\symb{list}}, \mapsTo{\prgvar{next}}{\symb{NULL}})}}
    \\
    {\scriptsize \mapsTo{\symb{d}}{\smalloperator{tv}\text{(void, \symb{?d})}}}
    \\
    {\scriptsize \mapsTo{\symb{length}}{\smalloperator{tv}\text{(int, \em{2})}}}
  }
 }
\\
\kallLarge{\mempc{}}{
{\scriptsize \symb{list} \neq \symb{NULL} \; \wedge}
\\
{\scriptsize \symb{list}\Rightarrow\prgvar{next} = \symb{NULL}}
}
\]
Since there are no observer paths returning different values %because
                                %there is only one path (the one for
                                %the computed pattern) 
and the associated return value is the integer 2, then the equation
$\tt length(list)=2$ is computed as a (partial) explanation for the
final pattern under consideration.
Thus, this term is added to the set of equations $eqSet$ that are
computed by  Algorithm~\ref{alg:explanations}.
\end{example}

A preliminary implementation of our  specification inference
methodology has been developed in the prototype system
\KingSpec{}, that is available at \href{safe-tools.dsic.upv.es/kindspec2}{\tt safe-tools.dsic.upv.es/kindspec2}.
The system is built upon a centralized call-and-return architecture that is shown in Figure \ref{fig:kindspec2}, where the main, front-end module orchestrates the different subprocesses of the inference algorithm and glues together their results. Given the source code of the program to be analyzed and the program function {\it f} whose specification needs to be inferred, first the system invokes the \K{} interpreter, {\tt krun}, providing it (through the coupling module) the name of the modifier function {\it f} and an initial empty state. By using the compiled definition of the programming language (\KernelC{}, in our case), the \K{} interpreter carries out the symbolic execution of {\it f}. As a result, a textual representation of the leaves of the symbolic tree is returned, and then parsed in order to obtain a higher level representation based on internal data structures and objects. Once the information retrieved from the final states of the execution is available, the explanation module synthesizes the specification as described in Algorithm \ref{alg:explanations}, and then   a conveniently simplified version of the axioms is output to the user. 
%Finally, the discovered specification is printed through the standard output to show it to the user.

% The specification ,that is based on %the 
% % framework 
% the new symbolic  \K{} infrastructure. %, called \KingSpec{}.
% %Differently from the implementation presented in \cite{AFV13}, the
% %implementation can easily be adapted to be applied to any language
% %with a \K{} definition.
\begin{figure}
\centering
\includegraphics[scale=0.52]{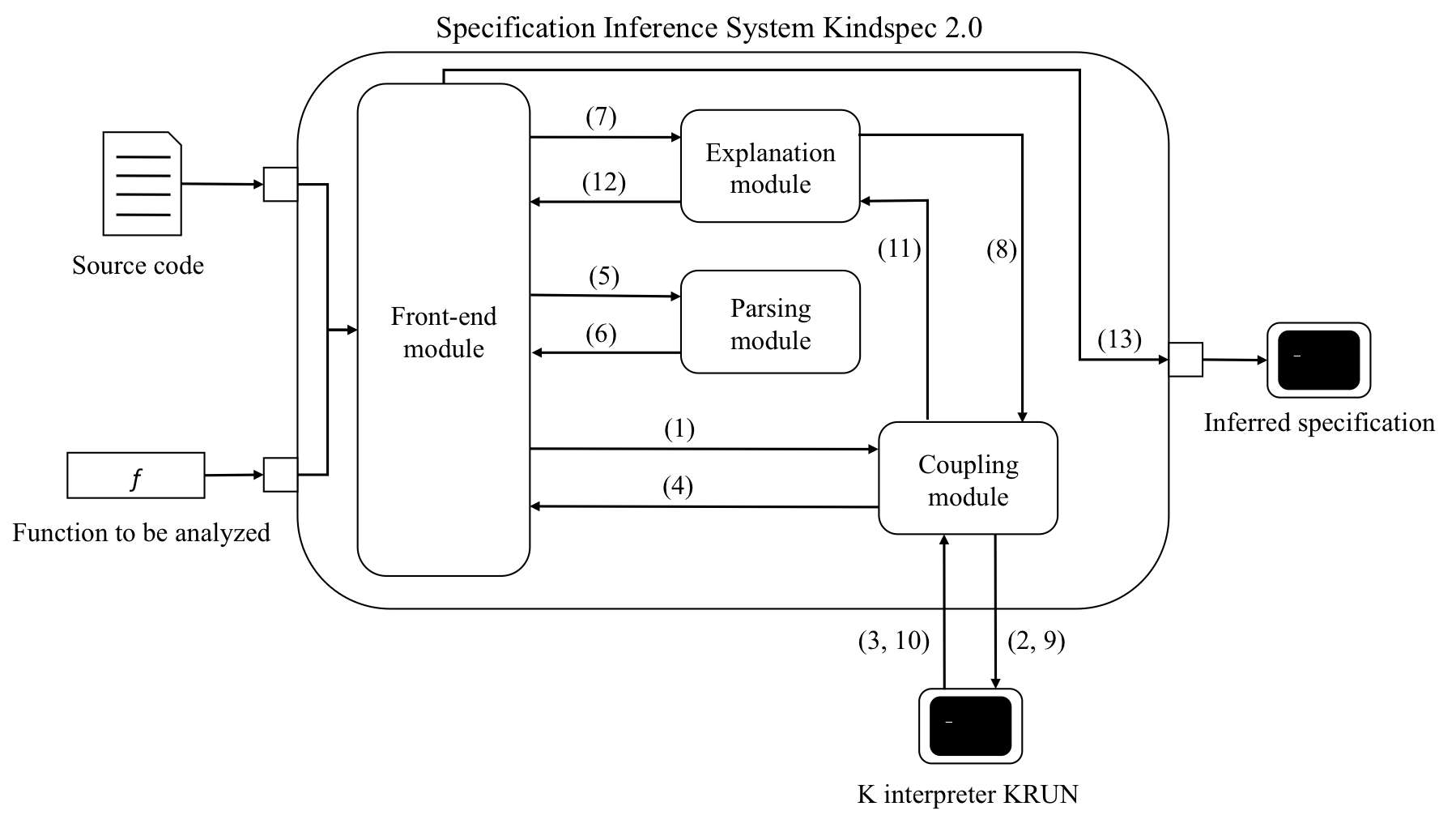}
\caption{Architecture of the inference system \KingSpec.}
\label{fig:kindspec2}
\end{figure}

{%\color{red}
The specification computed for our leading example is
  shown below: 
\begin{align*}
\begin{minipage}{.40\linewidth}
\small
\[\left(
\begin{array}{l}
\mathtt{length(list)}=\mathtt{0}\; \wedge\\
\mathtt{reverse(list)}=\mathtt{NULL}\; \wedge\\
\mathtt{find(list,data)}=\mathtt{0}\; \wedge\\
\mathtt{init(list)}=\mathtt{NULL}\; \wedge\\
\mathtt{last(list)}=\mathtt{NULL}
%\mathtt{remove}=\mathtt{NULL}
\end{array}
\right)\]
\end{minipage}
& \Rightarrow 
\begin{minipage}{.40\linewidth}
\small
\[\left(
\begin{array}{l}
\mathtt{length(list')}=1\; \wedge\\
\mathtt{reverse(list')}=\mathtt{list}\; \wedge\\
\mathtt{find(list',data)}=\mathtt{1}\; \wedge\\
%\mathtt{head(list')}=\mathtt{list'}\; \wedge\\
\mathtt{init(list')}=\mathtt{NULL}\;\wedge\\ %\mathtt{list}\; \wedge\\
\mathtt{last(list')}=\mathtt{data}\; \wedge\\
\mathtt{ret}=\mathtt{list'}
\end{array}
\right)\]
\end{minipage}
\\
\begin{minipage}{.40\linewidth}
\small
\[\left(
\begin{array}{l}
\mathtt{length(list)}=1
\end{array}
\right)\]
\end{minipage}
& \Rightarrow 
\begin{minipage}{.40\linewidth}
\small
\[\left(
\begin{array}{l}
\mathtt{length(list')}=2\; \wedge\\
\mathtt{find(list',data)}=\mathtt{1}\; \wedge\\
\mathtt{last(list')}=\mathtt{data}\; \wedge\\
\mathtt{ret}=\mathtt{list'}
\end{array}
\right)\]
\end{minipage}
\\
\begin{minipage}{.40\linewidth}
\small 
\[\left(
\begin{array}{l}
\mathtt{length(list)}=2
\end{array}
\right)\]
\end{minipage}
& \Rightarrow 
\begin{minipage}{.40\linewidth}
\small
\[\left(
\begin{array}{l}
\mathtt{length(list')}=3\; \wedge\\
\mathtt{find(list',data)}=\mathtt{1}\; \wedge\\
\mathtt{last(list')}=\mathtt{data}\; \wedge\\
\mathtt{ret}=\mathtt{list'}
\end{array}
\right)\]
\end{minipage}
\end{align*}
Note that, in contrast to the two axioms of Example~\ref{ex:specification}, three axioms are
  computed. This
  is due to the unrolling of loops. Note that the second and third
  computed axioms are instances of the second axiom of the intended
  specification.}

Similarly to \cite{AFV13}, due to bounded loop unrolling we cannot ensure completeness of the
inferred specifications since we do not cover all possible execution
paths. 
This is evident when comparing
%Note the difference of 
the automatically inferred %computed 
axioms shown in  the pattern above % (\ref{fig:ex8}) %in the
   w.r.t. the expected specification given in Example~\ref{ex:specification}. An effective generalization
 methodology is needed to properly  cover all possible executions without incurring (hopefully)  in significant loss of correctness.
 We informally discuss our key ideas towards this endeavour in the following subsection.

\subsection{Future directions}\label{sec:future}
%There are certainly many ways that our prototype implementation can be improved.

%This paper presents an ongoing work that, as such, has different
%tasks, both from the practical and from the theoretical point of view,
%to be attacked. 

We are currently working on defining a
\emph{generalization} algorithm that can  distill more general  axioms (such as the second axiom in
Example~\ref{ex:specification}) that we are not yet able to obtain. We follow the common synthesis approach that
is based on  using ``skeletons'' of generalizations, which are 
then refined to obtain a correct generalization of a set of axioms (w.r.t. the skeleton).
The function that computes such skeletons basically induces them from iterations (loops and recursive calls)
and is considered to be a parameter of the algorithm.
Without entering into too much detail, candidate skeletons are given by a so-called ``admissible template'', that is, a non-ground \K\ term that is used to guess 
  the form that a given general axiom can have.
A common drawback when resorting to skeletons is that the burden of defining/selecting the most suitable  
templates for a given problem usually rests with the user; hence usability is a key point that 
we cannot dismiss. 
Even if extensive research is still  needed, our preliminary
experiments reveal that axioms like the aforementioned more general one  can be easily inferred automatically. Obviously, since we are using a threshold to
stop loops,  correctness   cannot be ensured for all the general axioms that we compute, but
they can still be useful for other verification processes or  even   be 
verified afterwards.
A second, longer-term direction for research is to follow the abstraction-based, subsumption approach for
symbolic execution of \cite{APV2009-LazyInit} to finitize symbolic execution while getting rid of any thresholds.

% Nevertheless, when we consider loops that are characterized by an
% invariant that matches into a given set of invariant templates, then
% we could guarantee both the soundness and completeness of our
% technique.

%\alicia{lo de abajo habr\'{\i}a que decir (si queremos incluirlo) que es una posibilidad que necesita bastante m\'as tiempo que lo de arriba...}

%This basically consists of   introducing an abstract over-approximation of   states so that the
% path conditions in the final states
%of branches would contain abstracted variables that would encode the
%generalization. Obviously, t

From the experimental point of view, there are certainly several ways that our prototype implementation can be improved.
A refinement post-processing  was
defined in   \cite{AFV13} that  improves the quality of inferred specifications.
Roughly speaking, when an observed pattern  
cannot be explained because its symbolic
execution leads to final patterns that  do not agree in the same result,
the call pattern is (incrementally) split into multiple refined patterns 
until the considered observers eventually suffice to explain it.
We plan to implement this refinement process in \KingSpec{} and   measure the inference power   gains.
% Moreover, although the computed specifications are the same as in our previous
%work, we want to compare the new approach in terms of efficiency. 
Actually, the main motivation of our work was not  to improve
efficiency but rather to improve   robustness, generality and mantainability.
%  since \MatchC{} could not be adapted to deal with other
% programming languages different from \C.\

%% %----------------------------------SECTION-----------------------------------%
%\section{Conclusions and related Work}\label{sec:conc}
%
%\input{Conclusion.tex}

%--------------------------------BIBLIOGRAPHY--------------------------------%
\bibliographystyle{eptcs}
\bibliography{kspec-2015}

%\appendix
%\section{A case study of specification inference}\label{InferenceExample}
%
%\input{InferenceExample.tex}
\end{document}